\documentclass [preprint, superscriptaddress] {revtex4}
\usepackage{graphicx,latexsym,makeidx}
\begin{document}
\draft
\title {Magnetotransport Properties of Quasi-Free Standing Epitaxial Graphene Bilayer on
SiC: Evidence for Bernal Stacking}
\author {Kayoung Lee}
\affiliation {Microelectronics Research Center, The University of
Texas at Austin, Austin, TX 78758}
\author {Seyoung Kim}
\affiliation {Microelectronics Research Center, The University of
Texas at Austin, Austin, TX 78758}
\author {M. S. Points}
\affiliation {Microelectronics Research Center, The University of
Texas at Austin, Austin, TX 78758}
\author {T. E. Beechem}
\affiliation {Sandia National Laboratories, Albuquerque, NM 87185, USA}
\author {Taisuke Ohta}
\affiliation {Sandia National Laboratories, Albuquerque, NM 87185, USA}
\author {E. Tutuc}
\affiliation {Microelectronics Research Center, The University of
Texas at Austin, Austin, TX 78758}
\date{\today}
\begin{abstract}
We investigate the magnetotransport properties of quasi-free standing epitaxial graphene bilayer on SiC, grown by atmospheric pressure graphitization in Ar,
followed by H$_2$ intercalation. At the charge neutrality point the longitudinal resistance shows an insulating behavior, which follows a temperature dependence consistent with variable range hopping transport in a gapped state. In a perpendicular magnetic field, we observe quantum Hall states (QHSs) both at filling factors ($\nu$) multiple of four ($\nu=4, 8, 12$), as well as broken valley symmetry QHSs at $\nu=0$ and $\nu=6$. These results unambiguously show that the quasi-free standing graphene bilayer grown on the Si-face of SiC exhibits Bernal stacking.

\vspace{\baselineskip}	
\end{abstract}
\maketitle

Graphene bilayers in Bernal stacking \cite{novoselov_natphys06} exhibit a transverse electric field tunable band-gap \cite{mccannprl06,min}, as evidenced by angle-resolved photoemission \cite{ohta} and transport measurements \cite{castro07, oostinga}, a property that renders this material attractive for device applications.
Bernal stacking is the lowest energy structure, and is found in the natural graphite crystal.
Recent studies of graphene bilayer grown on SiC \cite{riedl} and Cu \cite{lee,yan}
have suggested the presence of Bernal stacking, based primarily on electron microscopy and Raman spectroscopy.
However, there has been no firm evidence of Bernal stacking based on electron transport, which in turn would allow
an assessment of its potential role for electronic devices \cite{xia}. It is therefore of high interest to explore the electron transport
in {\it grown} graphene bilayers, in order to determine the stacking of the layers and the key transport properties.

Here we investigate the transport properties of epitaxial graphene bilayer grown by
atmospheric pressure graphitization of SiC followed by H$_2$ intercalation, which renders
the graphene quasi-free standing. Using top gated Hall bars with Al$_2$O$_3$ dielectric
we probe the magneto-transport up to magnetic fields of 30 T and temperatures down to 0.3 K.
The devices show a high field-effect mobility of 2,600 - 4,400 cm$^2$/Vs, which changes little from room temperature down to 0.3 K,
as well as a strong insulating behavior near the charge neutrality point. The magneto-transport data reveal
quantum Hall states (QHSs) at filling factors $\nu=4, 8, 12$, consistent with the four-fold, spin and valley degenerate
Landau levels (LL) in Bernal stacked (A-B) graphene bilayer. More interestingly, the data also reveal
developing broken valley symmetry QHSs at filling factors $\nu=0$ and $\nu=6$, which testify to the
high sample quality. Further supported by Raman spectroscopy and low energy electron microscopy (LEEM) data, these results
unambiguously show that the quasi-free-standing epitaxial graphene bilayer grown on the Si face of SiC substrates
exhibits Bernal stacking.

The graphene bilayer films studied in this paper are produced via a two step process beginning with a starting substrate of 6H-SiC(0001) (Si-face, 2.1$\times$10$^{11}$ $\Omega \cdot$cm, II-VI Incorporated). Prior to graphitization, the substrate is hydrogen etched (45$\%$ H$_2$ - Ar mixture) at 1350 $^{\circ}$C to produce well-ordered atomic terraces of SiC. Subsequently, the SiC sample is heated to 1000 $^{\circ}$C in a 10 $\%$ H$_2$ - Ar mixture, and then further heated to 1550 $^{\circ}$C in an Ar atmosphere \cite{emtsev,pan}. This graphitization process results in the growth of an electrically active graphene layer on top of the buffer layer, covalently bound to the substrate. Finally, hydrogen intercalation was carried out using 45$\%$ H$_2$ - Ar mixture at 800 $^{\circ}$C \cite{riedl}, in order to decouple the buffer layer from the substrate. As we show here, the two graphene layers, a formally buffer layer (decoupled via hydrogen intercalation) and a monolayer graphene (formed via graphitization process), are in Bernal stacking.

Both the number of layers and the quality of the resulting material are probed by Raman spectroscopy acquired over a 25$\times$25 $\mu$m region, using a 532 nm excitation wavelength, 5 mW power, and 500 nm spot size.  Figure 1(a) shows a typical Raman spectrum of the graphene bilayer sample in which the contribution from SiC substrate has been subtracted. Well defined spectral features characteristic of graphene's G ($\sim$1600 cm$^{-1}$) and 2D-band ($\sim$2700 cm$^{-1}$) are observed.
The 2D-band was well fit utilizing four Lorentzian peaks each with a width of $\sim$35 cm$^{-1}$ confirming that the presence of graphene bilayer [Fig. 1(a) inset]. The Raman spectra indicate that the sample is of high quality as the intensity ratio of the defect induced D-band to G-band, while present is less than 0.1 for the great majority of the sample area [Fig. 1(b)]. The presence of graphene bilayer is further confirmed by the energy dependence spectrum of the specular electron reflection \cite{ohta2008}, and by the well ordered 6-fold low energy electron diffraction (LEED) pattern acquired with LEEM [Fig. 1(c)]. No turbostatic disorder was observed.

To probe the transport properties of these graphene bilayers we fabricate top-gated Hall bars.
The Hall bar location on the substrate is first chosen using optical and atomic force microscopy (AFM)
in order to identify an appropriately wide terrace [Fig. 1(d)]. The dashed
contour of Fig. 1(d) is an example of an 8 $\mu$m-wide terrace, onto which a Hall bar is
subsequently defined. Raman imaging confirms that there is only graphene bilayer within these terraces,
as panel (b) inset shows that the 2D mode is rather uniform inside terraces.
E-beam lithography and O$_2$ plasma etching are used to pattern the
Hall bar active area; the graphene is etched outside the Hall bar to prevent parallel current
flow. Metal contacts are realized using a second e-beam lithography step, followed by a 40 nm Ni
deposition and lift-off. To deposit the Al$_2$O$_3$ gate dielectric, we first deposit a 1.5 nm thick Al film.
The sample is then exposed to ambient, and transferred to an atomic layer deposition (ALD) chamber.
The ambient exposure causes the Al interfacial layer to fully oxidize \cite{dignam}, and
provides nucleation centers for the subsequent ALD process. A 15 nm thick Al$_2$O$_3$ top dielectric film is then
deposited using trimethylaluminum as Al source, and H$_2$O as an oxidizer \cite{kim}. A third
e-beam lithography step, followed by Ni deposition and lift-off are used to pattern the top gate
[Fig. 2(a) inset]. The corresponding dielectric capacitance for this stack is 245 nF/cm$^{2}$,
with an average dielectric constant of $k=4.6$.

Four point longitudinal ($\rho_{xx}$) and Hall ($\rho_{xy}$) resistivity measurements are performed
at temperatures ($T$) down to 0.3 K, and magnetic fields ($B$) up to 30 T, using low-current,
low-frequency lock-in techniques. The carrier density ($n$), and its dependence on the top gate voltage ($V_{TG}$)
are determined from Hall measurements, as well as the filling factors of the quantum Hall states observed in high magnetic fields.
The longitudinal ($\sigma_{xx}$) and Hall ($\sigma_{xy}$) conductivities are determined via a tensor inversion from the measured
resistivities.

In Fig. 2(a) we show $\sigma_{xx}$ vs. $V_{TG}$, measured at $T=290$ K and 0.6 K, and at $B$ = 0 T, revealing an
ambipolar characteristic with a charge neutrality point at a positive $V_{TG}$ value. Away from the charge neutrality point, the $\sigma_{xx}$
vs. $V_{TG}$ data show a linear dependence up to the highest $V_{TG}$ values, with a corresponding field-effect mobility of 2,600 - 4,400 cm$^2$/Vs at $T$=290 K \cite{mobility}. At $T=0.6$ K, where the measurement was performed in a wider $V_{TG}$ range, the $\sigma_{xx}$ vs. $V_{TG}$ linear dependence persists
down to -4 V, for a gate voltage overdrive of up to 5 V. The linear $\sigma_{xx}$ vs. $V_{TG}$ dependence contrasts data reported in mono-layer graphene,
where neutral impurity scattering, which is density independent, limits the conductivity at high $V_{TG}$ values.
In contrast, the neutral impurity scattering remains proportional to $n$ in graphene bilayers \cite{adam08}.

Figure 2(b) shows $\rho_{xx}$ vs. $V_{TG}$ at different $T$ values, measured at $B=0$ T. Remarkably, $\rho_{xx}$ measured
near the charge neutrality point is strongly temperature dependent, with an insulating behavior. The insulating phase at
the charge neutrality point suggests an energy gap, and contrasts data from mono-layer graphene with comparable mobility,
where $\rho_{xx}$ at the charge neutrality point is weakly temperature dependent. In contrast, graphene bilayers in the presence
of an applied transverse electric field open a tunable energy gap between the conduction and valence bands, thanks to the layer on-site energy asymmetry \cite{mccannprl06}. We posit the presence of a transverse electric field in our samples, due to unintentional doping and
the asymmetry of the device structure. The inset of Fig. 2(b) shows the $T$-dependence of the resistivity measured at the charge neutrality point ($\rho_{NP}$).
The data clearly follow a $\propto e^{(T_0/T)^{1/3}}$ dependence for $T$ lower than $100$ K, indicating that variable range hopping
rather than thermally activated conduction controls the electron transport at low temperatures.
This has been attributed to disorder-induced localized states in the gap, which reduce the effective energy gap \cite{zou,taychatanapat},
and render the $T$-dependence of the $\rho_{NP}$ weaker than the exponential $\propto e^{\Delta/2k_B T}$ dependence expected for a band insulator with
an energy gap $\Delta$; $k_B$ is the Boltzmann constant. The extracted $T_0$ value corresponding to Fig. 2(b) inset data is 0.6 K, similar
to previously reported values on exfoliated graphene bilayers \cite{zou,oostinga}. A fit of the $\rho_{NP}$ vs. $T$ data of Fig. 2(b) inset to
the exponential $\propto e^{\Delta/2k_B T}$ dependence for $T\geq120$ K, yields an energy gap at the charge neutrality point of $\Delta=20\pm6$ meV.

While Fig. 2 data at the charge neutrality point show an energy gap, suggestive of a graphene bilayer,
the most important finding of this study is presented in Fig. 3. Figure 3(a) shows $\rho_{xx}$ and $\rho_{xy}$ vs. $V_{TG}$,
measured at a high magnetic field of $B=30$ T, and a temperature $T=0.3$ K.
The data reveal strong QHSs at integer filling factors $\nu=4$, and $\nu=8$, marked by vanishing or local minima in $\rho_{xx}$,
and corresponding $\rho_{xy}$ plateaus. The QHSs at integer filling factors multiple of four in Fig. 3 stem from the four-fold degeneracy associated with
spin and valley of each Landau level, and unambiguously identify the material as being a Bernal stacked (A-B) graphene bilayer.
In contrast, thanks to the $\pi$-Berry phase of chiral Dirac fermions, the strongest QHSs in mono-layer graphene are present at integer fillings
$\nu=2$, 6, 10, etc... \cite{novoselov05,zhang05} Figure 3(b) data show $\sigma_{xy}$ vs. $V_{TG}$ at different values of the $B$-field.
The $\sigma_{xy}$ plateaus at $\nu\cdot e^2/h$ values, which become stronger with increasing the $B$-field confirm
the presence of QHSs at fillings $\nu=4$, 8, 12 consistent with a Bernal stacked graphene bilayer. More interestingly, Fig. 3 data reveal a strong
QHS at $\nu=0$, and a developing QHS at $\nu=6$. While a possible explanation for the $\nu=6$ QHS is a mixture of graphene mono- and bilayer,
this is ruled out by the absence of the other QHSs associated with mono-layer graphene, in particular the $\nu=2$ QHS. The absence of graphene monolayers
is further corroborated by LEEM measurements. On the other hand, the presence of $\nu=0$ and $\nu=6$ QHSs is fully consistent with a graphene bilayer in the presence of a transverse electric field, a finding which is also consistent with Fig. 2 data. We expand below on this argument.

At zero transverse electric field ($E$), an eight-fold degenerate Landau level (LL) is located at zero energy (charge neutrality point);
it consists of the $N$ = 0, 1 LLs with their respective valley and spin degeneracies. The $N=0$, 1 LLs are layer polarized
and split in the presence of a transverse $E$-field depending on the on-site energy of the layer (valley) degree of
freedom \cite{mccannprl06}, leading to a QHS at $\nu=0$. Similarly, the presence of a transverse $E$-field
breaks the valley degeneracy of $|N|\geq2$ LLs \cite{nakamura}, but the corresponding energy splitting is smaller than that of $N=0$, 1 LLs.
This explains the absence of $\nu=6$ QHS in exfoliated graphene bilayers on SiO$_2$, where the LL disorder broadening can obscure
the $E$-field induced splitting. In that regard, the observation of a $\nu=6$ QHS in quasi-free standing graphene bilayers
is interesting, and testifies to a reduced disorder and LL broadening in these samples, by comparison to dual-gated graphene bilayers
on SiO$_2$ substrates, and with a similar top-gate stack \cite{kim11}. We note that back-gated graphene bilayers on SiO$_2$
samples can exhibit higher mobilities \cite{zhao}.

In the remainder, we explore further the $\nu=0$ QHS in these samples. Depending on the transverse $E$-field, the $\nu=0$ QHS in
graphene bilayers can be either spin-polarized at small $E$-fields, or valley- (layer) polarized at large $E$-fields \cite{gorbar10,toke10}.
If the $E$-field is varied at a given perpendicular magnetic field, the $\nu=0$ QHS undergoes a transition from spin-to-valley-polarized at a critical $E$-field ($E_c$).
Experimental studies \cite{weitz,kim11} on dual-gated A-B (Bernal) graphene bilayers exfoliated from natural graphite show a linear dependence of $E_c$
on $B$, with a slope of 12-18 (mV/nm)$\cdot$T$^{-1}$. Figure 4(a) shows $\rho_{xx}$ vs. $V_{TG}$, measured at $B$ = 30 T, and at different temperatures.
The data reveal an insulating phase near the charge neutrality point, consistent with a LL splitting at zero energy
responsible for the presence of the $\nu=0$ QHS of Fig. 3(b). The inset of Fig. 4(a) shows the Arrhenius plot of $\rho_{xx}$ vs. $T^{-1}$ at $\nu=4$ QHS;
the data follow an activated $T$-dependence, $\rho_{xx}\propto e^{-\Delta_4 / 2 k_B T}$ with
an energy gap $\Delta_{4}=16$ K for $\nu=4$. Figure 4(b) shows $\rho_{xx}$ vs. $V_{TG}$ measured at $T=0.3$ K and
at different values of the $B$-field. Figure 4(b) data show a weak dependence of $\rho_{xx}$ on $B$
at $\nu=0$. This finding, analyzed by comparison with existing data in exfoliated A-B bilayers \cite{kim11} suggest
that the $\nu=0$ is neither spin-polarized or valley-polarized in the range of $B$-fields probed here, but rather in the vicinity of the spin-to-valley-polarized transition (see supplementary material). Indeed, if the $\nu=0$ QHS was spin polarized, we would expect $\rho_{xx}$ at $\nu=0$ to increase with $B$.  Conversely,
if the $\nu=0$ QHS was valley polarized a decrease of $\rho_{xx}$ with the $B$-field is expected. This argument provides us with
an estimate of the transverse $E$-field across the bilayer of 0.35$\pm$0.07 V/nm.

In summary, we investigate the magnetotransport in quasi-free-standing graphene bilayers on SiC. We observe QHSs at
fillings $\nu=0, 4, 6, 8, 12$, consistent with a Bernal stacked graphene bilayer in the presence of a transverse field.
An insulating state observed at $B=0$ T, near the charge neutrality point indicates the opening of an energy gap, in agreement
with the expected response of a Bernal stacked graphene bilayer. These findings corroborated with earlier microscopy
studies \cite{riedl} unambiguously identify the Bernal stacking arrangement of graphene bilayers on the Si-face of SiC substrates,
and render this system particularly attractive for electronic and optoelectronic device applications, thanks to its high mobility,
tunable energy gap, and high on/off ratio.

The work at University of Texas at Austin was supported by NRI, DARPA, NSF (DMR-0819860), and the NINE program.
Part of this work was performed at the National High Magnetic Field Laboratory, which is supported by NSF (DMR-0654118),
the State of Florida, and the DOE. The work at Sandia Labs was supported by LDRD, and performed in part at CINT, a US DOE, Office of Basic Energy Sciences user facility (DE-AC04-94AL85000). Sandia is a multiprogram laboratory operated by Sandia Corporation, a Lockheed Martin company, for the US DOE's
National Nuclear Security Administration under contract DE-AC04-94AL85000. We are grateful to Guild Copeland and Anthony McDonald for
sample preparation and characterization, partly supported by the US DOE Office of Basic Energy Sciences, Division of Materials Science and Engineering.

\newpage

\begin{figure*}
\centering
\includegraphics[scale=1]{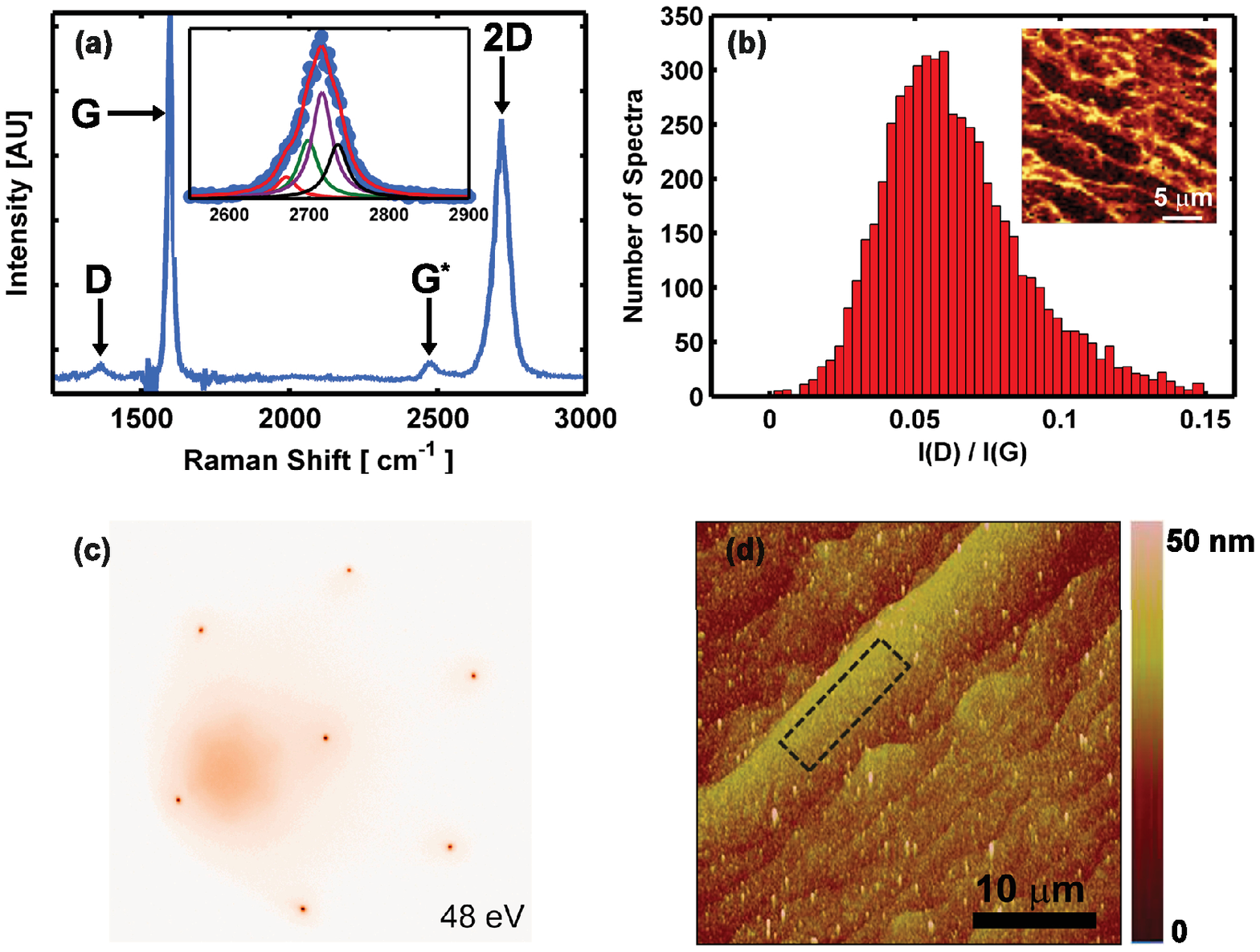}
\caption {\small{(color online) Raman, LEED and AFM characterization of the graphene bilayer sample. (a) Representative spectrum of the bilayer region, with the SiC response removed. The inset shows a 2D band fitted well using four Lorentzian functions, an indication of the graphene bilayer presence.
(b) Histogram of the I(D)/I(G) ratio acquired from Raman mapping (25$\times$25 $\mu$m$^{2}$, 75$\times$75 data points). The inset shows the spatial distribution of the total 2D peak width indicating that the bilayer is present on the terraces (dark regions within inset, total width $\simeq$50 cm$^{-1}$), whereas thicker graphene layers (bright regions of inset, total width $\simeq$70 cm$^{-1}$) are located at atomic steps originating from the SiC. (c) LEED pattern of a bilayer graphene obtained at the illuminating electron energy of 48 eV. (d) AFM topography of the graphene sample. The dashed contour on the plateau indicates the region used for device fabrication.}}
\end{figure*}

\begin{figure}
\centering
\includegraphics[scale=0.8]{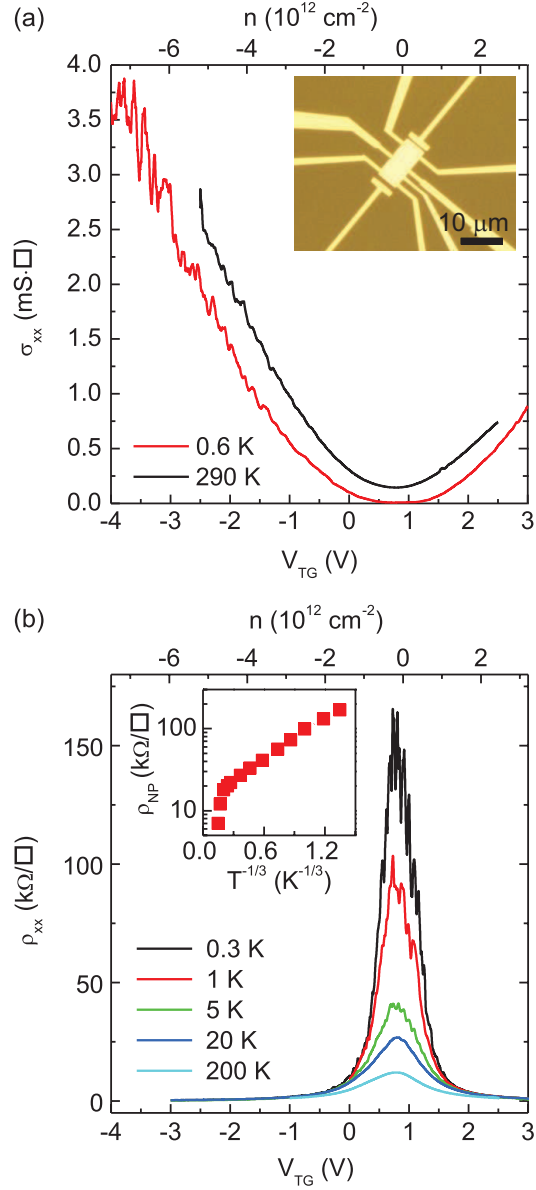}
\caption {\small{(color online) (a) $\sigma_{xx}$ vs. $V_{TG}$ measured at $T=290$ K and $T=0.6$ K. The top axis represents the carrier density ($n$);
positive values correspond to $n$-type carriers (electrons), while negative values correspond to $p$-type carriers (holes).
The inset shows an optical micrograph of the top-gated Hall bar. (b) $\rho_{xx}$ vs. $V_{TG}$ measured at different $T$ values.
The resistivity near the charge neutrality point shows a clear insulating behavior. The inset shows $\rho_{NP}$ vs. $T^{-1/3}$ on
a log-lin scale. The data are consistent with variable range hopping transport, measured in exfoliated bilayers
when a band gap opens as a result of an applied transverse electric field.}}
\end{figure}

\begin{figure}
\centering
\includegraphics[scale=0.8]{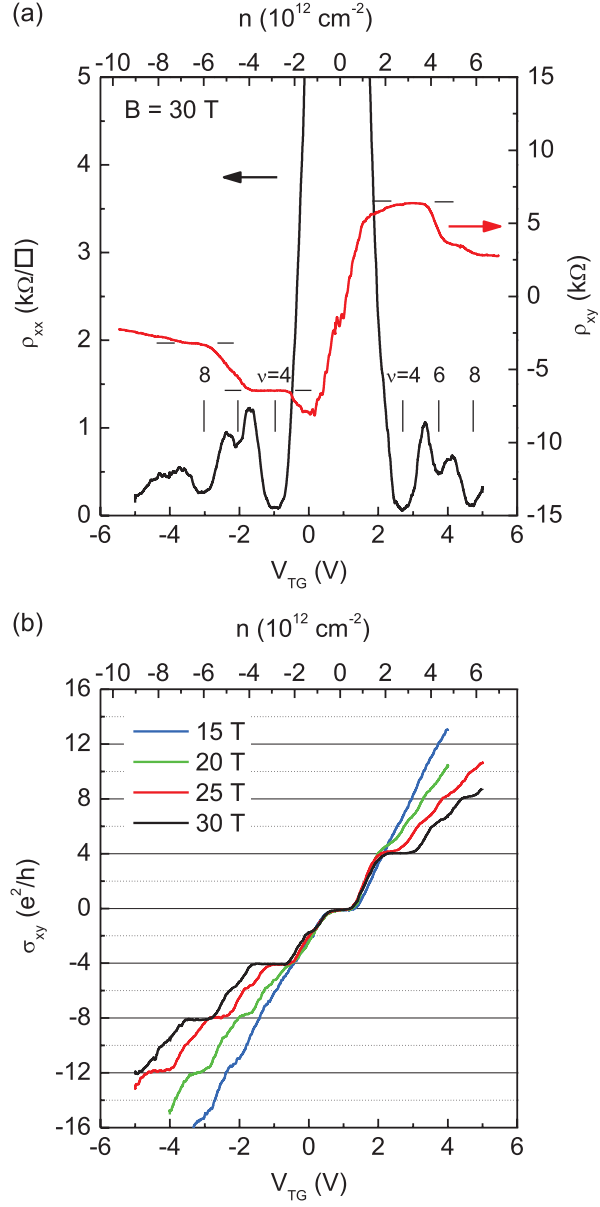}
\caption {\small{(color online) (a) $\rho_{xx}$ and $\rho_{xy}$ vs. $V_{TG}$ (bottom axis), and $n$ (top axis),
measured at $B=30$ T, and $T=0.3$ K. (b) $\sigma_{xy}$ vs. $V_{TG}$ (bottom axis) and $n$ (top axis) measured
at $T=0.3$ K, and different $B$-field values. The data show developing QHSs with increasing the $B$-field.}}
\end{figure}

\begin{figure}
\centering
\includegraphics[scale=0.8]{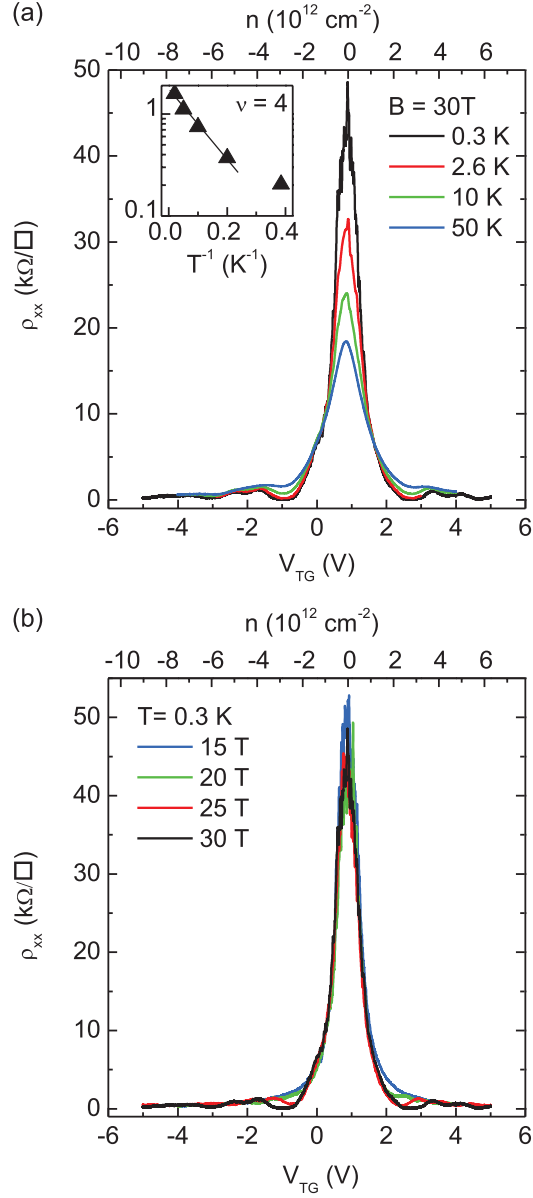}
\caption {\small{(color online) (a) $\rho_{xx}$ vs. $V_{TG}$ (bottom axis) and $n$ (top axis), measured for different $T$ values, at $B=30$ T.
Inset: $\rho_{xx}$ vs. $T^{-1}$ at $\nu=4$ on a log-lin scale. The $\nu=4$ energy gap is $\Delta_4=16$ K. (b) $\rho_{xx}$ vs. $V_{TG}$ at different
$B$-field values measured at $T=0.3$ K. Concomitantly with developing QHSs at increasingly higher $B$-field values, the $\rho_{xx}$ measured
at $\nu=0$ changes little with the $B$-field, suggesting the $\nu=0$ is not strongly spin or valley polarized in the range of $B$-fields explored here.}}
\end{figure}

\end{document}